\documentclass[a4paper,11pt]{article}
\usepackage{pos}

\title{The cosmic-ray sea explains the Galactic $\gamma$-ray and $\nu$ diffuse emissions from GeV to PeV}
 \ShortTitle{Galactic $\gamma$-ray and $\nu$ emissions from Cosmic Rays}

\author[a]{Pedro~De~La~Torre~Luque}

\author[b]{Daniele~Gaggero}

\author*[b]{Dario~Grasso}

\author[c,d,e]{Antonio~Marinelli}

\author[f]{Manuel~Rocamora}

\affiliation[a]{Instituto de Física Teórica, IFT UAM-CSIC, Departamento de Física Teórica,\\
Universidad Autónoma de Madrid, ES-28049 Madrid, Spain}

\affiliation[b]{INFN Sezione di Pisa, Polo Fibonacci, Largo B. Pontecorvo 3, 56127 Pisa, Italy}

\affiliation[c]{Dipartimento di Fisica ``Ettore Pancini'', Università degli studi di Napoli ``Federico II'', \\ 
Complesso Univ. Monte S. Angelo, I-80126 Napoli, Italy}

\affiliation[d] {INFN Sezione di Napoli, Complesso Univ. Monte S. Angelo, I-80126 Napoli, Italy}

\affiliation[e]{INAF -- Osservatorio Astronomico di Capodimonte, Salita Moiariello 16, I-80131, Napoli, Italy}

\affiliation[f]{Universit\"at Innsbruck, Institut für Astro- und Teilchenphysik, Technikerstr. 25/8, 6020 Innsbruck, Austria}

\emailAdd{dario.grasso@pi.infn.it}

\abstract{The LHAASO collaboration has recently released the spectrum and the angular
distribution of the gamma-ray Galactic diffuse emission from 1 TeV to 1 PeV measured with the
Kilometer-2 Array (KM2A) and the Water Cherenkov Detector Array (WCDA). We show that
those data are in remarkably good agreement with a set of pre-existing models that assume the emission
to be produced by the Galactic population of cosmic rays if its spectral shape traces that measured by CALET, DAMPE as well as KASCADE at higher energies. No extra-component besides the CR sea is needed to explain LHAASO results. 
Spatial dependent CR transport models, although not required to reproduce LHAASO results, are in better agreement with them respect to conventional ones and needed to consistently reproduce Fermi-LAT and neutrino data. }

\ConferenceLogo{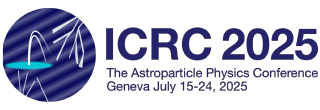}

\FullConference{39th International Cosmic Ray Conference (ICRC2025)\\
 15–24 July 2025\\
Geneva, Switzerland\\}

\begin{document}
\maketitle

\section{Introduction}

The Large High Altitude Air Shower Observatory (LHAASO) is a large area, wide field-of-view observatory for CRs and $\gamma$-rays ~\cite{LHAASO:2019qtb}.
This experiment combines measurements taken by the Kilometer-2 Array (KM2A) charged particle detector with those of the Water Cherenkov Detector Array (WCDA) allowing to cover an energy interval going from few TeV up to PeV.  
LHAASO is presently the best instrument suited to study the $\gamma$-ray Galactic Diffuse Emission (GDE) above the TeV. 
In 2023 the LHAASO collaboration released its measurement of the average diffuse flux in two different regions covering the Galactic Plane (GP) -- the inner ($15^{\circ} < l < 125^{\circ}$) and outer ($125^{\circ} < l < 235^{\circ}$) regions (both with $|b| < 5^{\circ}$) -- from $10$~TeV to $1$~PeV obtained with the KM2A array, in Ref.~\cite{LHAASO:2023gne}. 
More recently the collaboration updated those results including also the $\gamma$-GDE flux measurement performed with the WCDA from $1$ to $25$~TeV, allowing the release of the wide-band spectral distribution from 1 TeV to 1 PeV~\cite{LHAASO:2024lnz} hence almost connecting to Fermi-LAT results \cite{Fermi-LAT:2016zaq}.
Moreover LHAASO measured the latitude and longitude GDE profiles in the $10$-$63$~TeV and $63$-$1000$ TeV energy bands. 
The detection significance measured by the combined WCDA$+$KM2A was found to be 24.6 $\sigma$ and 9.1 $\sigma$ in the inner and outer regions respectively. 

It is important to point-out that in order to suppress the contamination of resolved (especially extended ones) and unresolved sources, LHAASO used a mask which cuts away a large fraction the low Galactic latitude ($\vert b \vert < 2^{\circ}$) region of the inner GP. 
This is different from the approach followed by the Tibet-AS$\gamma$ \cite{TibetASgamma:2021tpz} collaboration -- which also measured the GDE though only between 100 TeV and 1 PeV and with a smaller significance -- which masked only a small area extending $0.5^{\circ}$ around the nominal position of each known point like source. 
We notice that the GDE found by Tibet-AS$\gamma$ is significantly higher than LHAASO, the two measurements being in better agreement only if the latter were determined using no mask \cite{Kato:2025gva}. 
Even taking into account the possible enhancement and hardening of the GDE in the inner GP (see below),
this discrepancy suggests that Tibet signal may be partially contaminated by sources, especially extended ones (see also \cite{LHAASO:2023gne}).

LHAASO results represent an important achievement for astroparticle physics.
In fact, the GDE is a background for point like and, more critically, extended source observations as well as for dark matter indirect search. 
More importantly in the present context, its measurement provides an unique handle to determine the spectrum and the large scale spatial distribution of the Galactic Cosmic Ray (CR) population while direct detection of charged particles, with satellite and ground detectors, only probe relatively local CR properties. 
Moreover, at energies above few hundred TeV these properties can only be measured indirectly studying the secondary particle showers they produce in the atmosphere. That procedure introduces large systematics mostly connected to the simulation of the air showers development.  
In this regard we observe the large offset of the CR proton spectrum measured by KASCADE \cite{Apel:2013uni} respect to that published by IceTop \cite{IceCube:2019hmk} and, very recently, by the LHAASO collaboration \cite{LHAASO:2025byy}.
Waiting for experimental groups to get to a better agreement, the newly born PeV $\gamma$-ray astronomy offers an independent insight especially if exploited in combination with high energy neutrino astronomy so to get rid of the degeneracy between leptonic and hadronic $\gamma$-ray emissions. 

We stress, however, that even precisely knowing the CR spectra at the Earth position, is not sufficient to allow a realistic modelling of the GDE which is produced by the interaction of the CR hadrons and electrons with the interstellar gas and radiation field respectively.
In fact, due to not uniform spatial distribution of CR sources and to the shape of the Galaxy, also the CR large scale distribution is not expected to be uniform but rather peaked around the inner molecular ring, 
a couple of kpc away from the Galactic Centre (GC).
Furthermore, observational \cite{Gaggero:2014xla,Fermi-LAT:2016zaq} and theoretical arguments \cite{Cerri:2017joy} suggest that CR transport may also be spatial dependent giving rise to a hardening of the CR spectrum towards the GC.
Accounting for the -- also quite peaked towards the GC -- gas distribution all this can have a major impact on the GDE at energies above the TeV. 
In order to account for these, as well as other, effects so to produce as much as possible realistic and accurate emission maps to be used as templates by the current high resolution experiments, dedicated numerical codes have to be used. 
Here we use the DRAGON code \cite{Evoli:2016xgn} which in combination with HERMES \cite{Dundovic:2021ryb} allows to model the $\gamma$ and $\nu$ GDEs under the condition it reproduces -- as a working hypothesis to be tested at posteriori -- the primary and secondary CR spectra measured at the Earth.  High energy $\gamma$-ray attenuation due to the $\gamma-\gamma$ scatteting on the CMB is properly taken into account.

\begin{figure}[t]
\centering
\includegraphics[width=0.495\textwidth, height = 0.22\textheight]{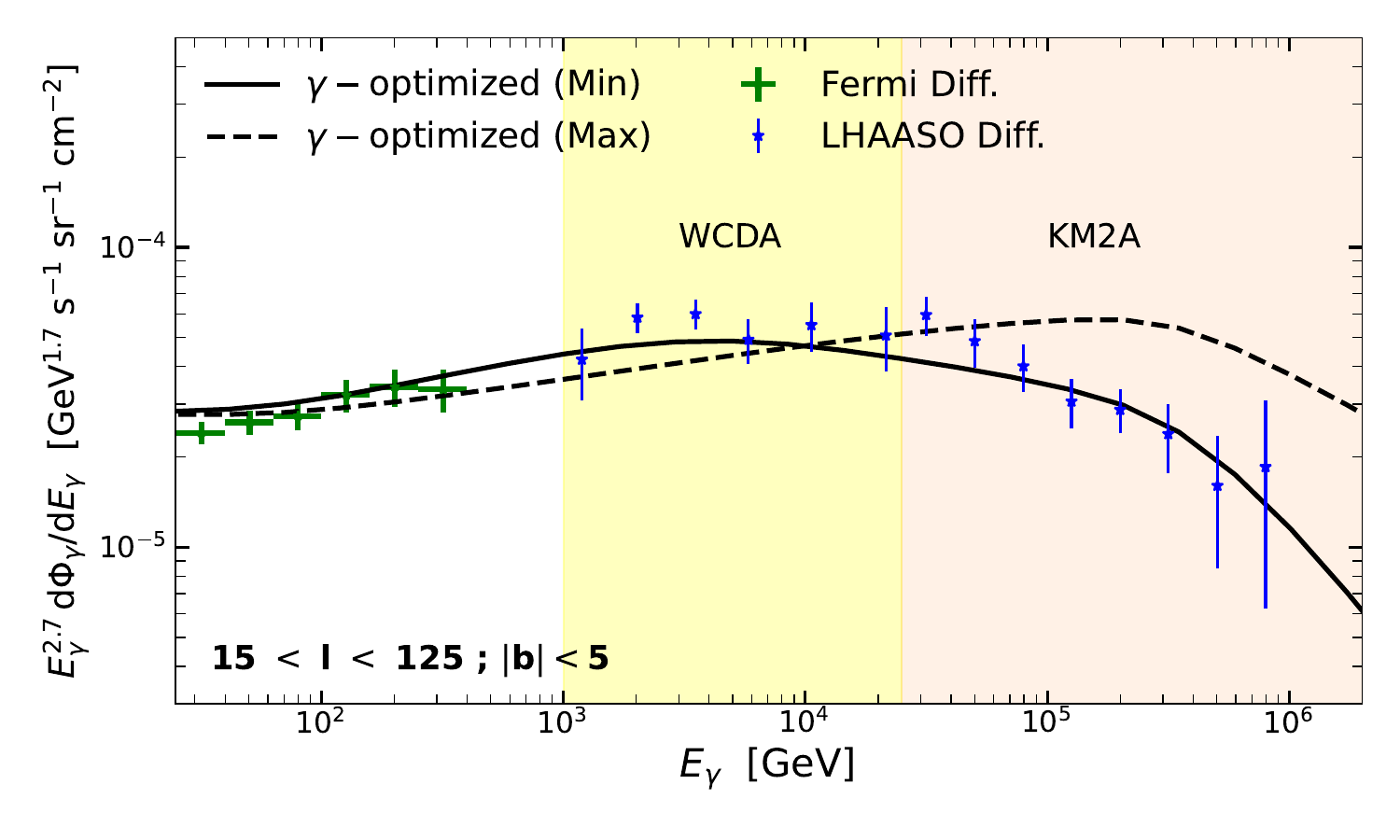}
\includegraphics[width=0.495\textwidth, height = 0.22\textheight]{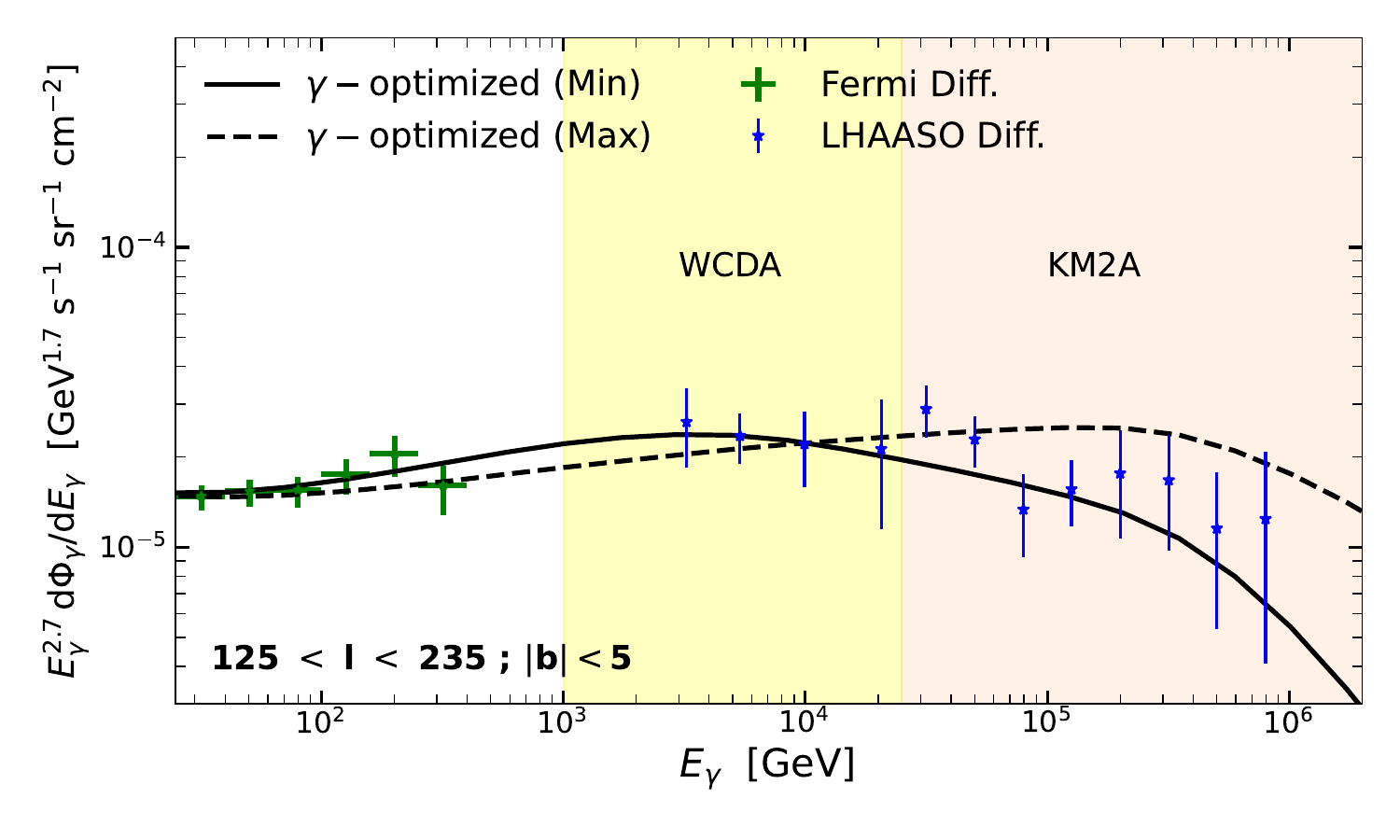}
\caption{The Galactic $\gamma$-ray diffuse emission spectra computed with the $\gamma$-optimized Min and Max models \cite{Luque:2022buq} 
are compared with  LHAASO (KM2A + WCDA) \cite{LHAASO:2024lnz} and Fermi-LAT data in the inner and outer Galactic Plane regions (see also \cite{DeLaTorreLuque:2025zsv}).
We shaded the energy region covered by the WCDA and KM2A in different colors, for clarity. We obtained Fermi-LAT diffuse data points applying the LHAASO mask and subtracting contribution from sources (4FGL catalog) and the EGB. Note as the models do not represent a fit of LHAASO data rather they were determined on the basis of CR and lower energy $\gamma$-ray data.  The curves would not change significantly replacing the models with the corresponding Base (space independent) ones for the reason explained in Sec.\ref{sec:models}.}
\label{fig:LHAASO_optimized_Comp}
\end{figure}

\begin{figure}[t]
\centering
\includegraphics[width=0.9\textwidth]{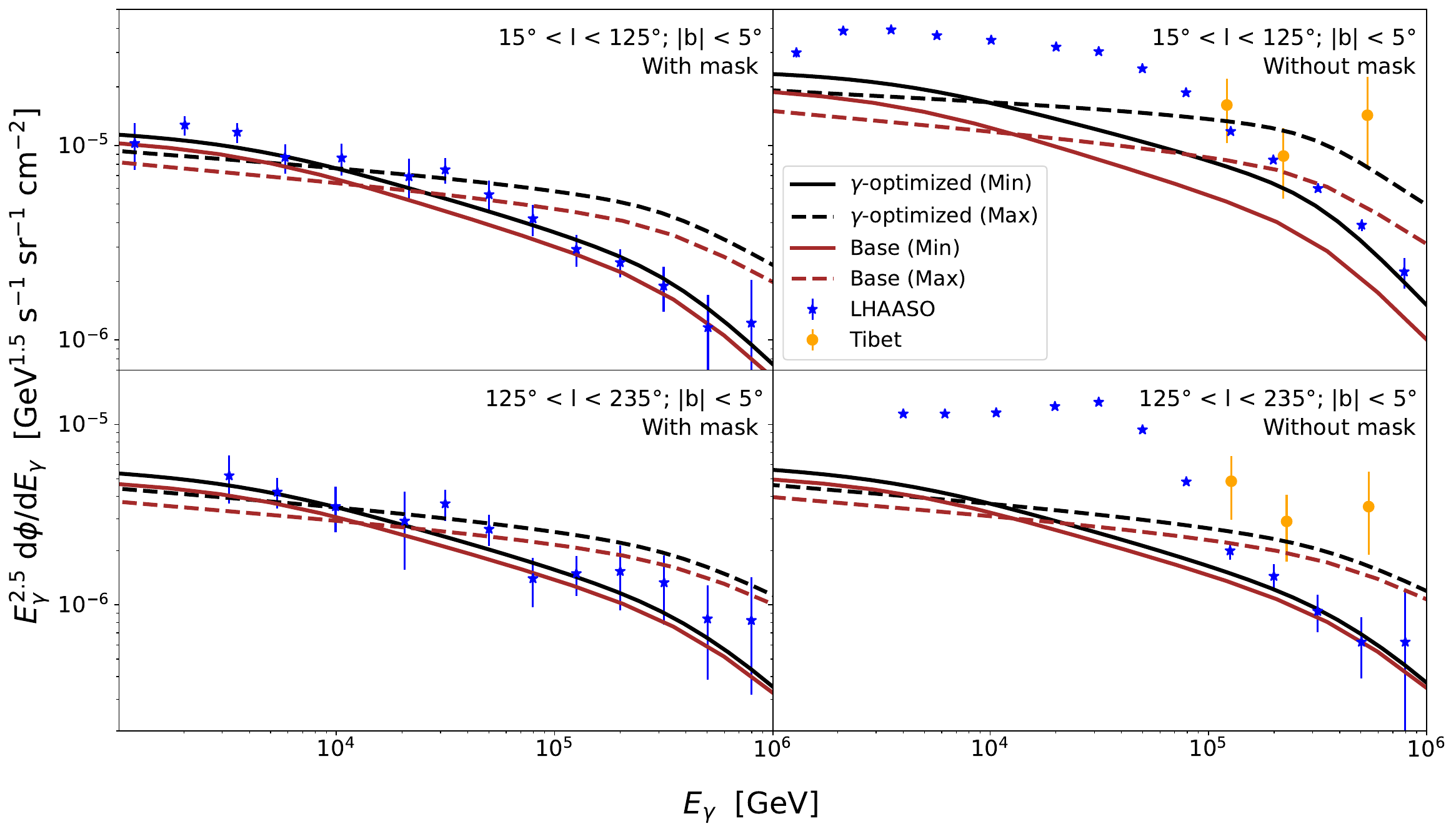}
\caption{We show here the GDE spectra computed with (left panels) and without (right panels) the mask used by LHAASO in the inner and outer regions probed by that experiment.
In the first case the models are compared with LHAASO KM2A and WCDA data. In the second with not-masked LHAASO and Tibet data. In the outer region Tibet data -- taken in a slightly different region -- are rescaled for illustrative reasons.}
\label{fig:Full_LHAASO_Comp}
\end{figure}

In Fig.~\ref{fig:LHAASO_optimized_Comp} we present (see also \cite{DeLaTorreLuque:2025zsv}) the comparison of two reference DRAGON models -- $\gamma$-optimized Min and Max -- with the GDE measured by LHAASO and Fermi-LAT \cite{Fermi-LAT:2012edv} data (extracted using the FERMITOOLS v.2.0.8) in the inner and outer regions probed by LHAASO.  For consistency, the same mask used by LHAASO was applied also to the Fermi-LAT data as well as to the model fluxes. 
Clearly, due to the mask cutting a large fraction of the inner GP where the GDE is stronger, the average fluxes showed in Fig.~\ref{fig:LHAASO_optimized_Comp} -- especially in the left panel -- are lower than those expected from the whole corresponding sky windows (see Fig.~\ref{fig:Full_LHAASO_Comp}).
The Min and Max setups were introduced to bracket the uncertainty in the shape of the source proton and helium spectra -- heavier nuclei giving a negligible contribution -- due to the scattered experimental air shower results mentioned in the above. 
The models, already presented in \cite{Luque:2022buq}, represent an upgrade (using the same CR propagation parameters) of the space-dependent KRA$_\gamma$ models \cite{Gaggero:2014xla}  which were shown to consistently reproduce the GDE measured by Fermi-LAT and several experimental data above the TeV \cite{Gaggero:2015xza}. 
In Ref.~\cite{Gaggero:2015xza} it was also shown that the KRA$_\gamma$ model predicts an enhanced high energy neutrino emission of the Galaxy with respect to conventional models. That prediction was confirmed in the IceCube collaboration 
discovery paper of the $\nu$-GDE where those models were used as templates in the fitting analysis \cite{IceCube:2023ame} (see Sec.\ref{sec:neutrinos}).

The remarkable agreement between the $\gamma$-optimized Min model and the experimental data over more than four energy decades is the main result presented in this contribution (see also Ref.~\cite{DeLaTorreLuque:2025zsv}).
This finding strongly support the scenario in which the $\gamma$-ray GDE is originated by the hadronic component of the Galactic CR sea.  No extra CR source component is required even for conventional transport models.
Moreover, it also implies that in the sky window considered by LHAASO -- masking a large fraction of the GP -- resolved and unresolved source can only give a minor contribution (see also \cite{Lipari:2024pzo} where we also showed that Inverse Compton emission is negligible in the energy band probed by LHAASO). 

Last but not least, we found a more than $10\sigma$ evidence that the Max model is incompatible with LHAASO $\gamma$-ray data
(the reduced $\chi^2$ is  $0.8$ for the Min model and 9.5 for the Max model, in the inner region, and $0.6$ and $1.9$ for the Min and Max models, respectively, in the outer region). 
This would be at odds with the CR proton spectrum measured by IceTop \cite{IceCube:2019hmk} and more recently by LHAASO \cite{LHAASO:2025byy} if its shape were representative of the local CR population and might require for a revision of current CR models  (see also \cite{Castro:2025wgf}).

\section{Conventional versus space dependent CR transport}\label{sec:models}

In this section we discuss to which extent LHAASO results may support spatial dependent CR transport models respect to conventional ones besides the evidences already provided by Fermi-LAT and IceCube (see Sec. \ref{sec:neutrinos}).

First of all we recall that the LHAASO-WCDA measured different GDE spectra in the inner ($15^{\circ} < l < 125^{\circ}$) and outer ($125^{\circ} < l < 235^{\circ}$) regions finding them to follow power-laws with indexes
$-2.67 \pm 0.05{~\rm stat}$ and $-2.83 \pm 0.19{~\rm stat}$  respectively \cite{LHAASO:2024lnz}. This provides a mild hint (with a significance of $\sim2\sigma$) that the primary CR spectrum may get harder toward the GC. 

\begin{figure}[t]
\centering
\includegraphics[width=0.6\textwidth]{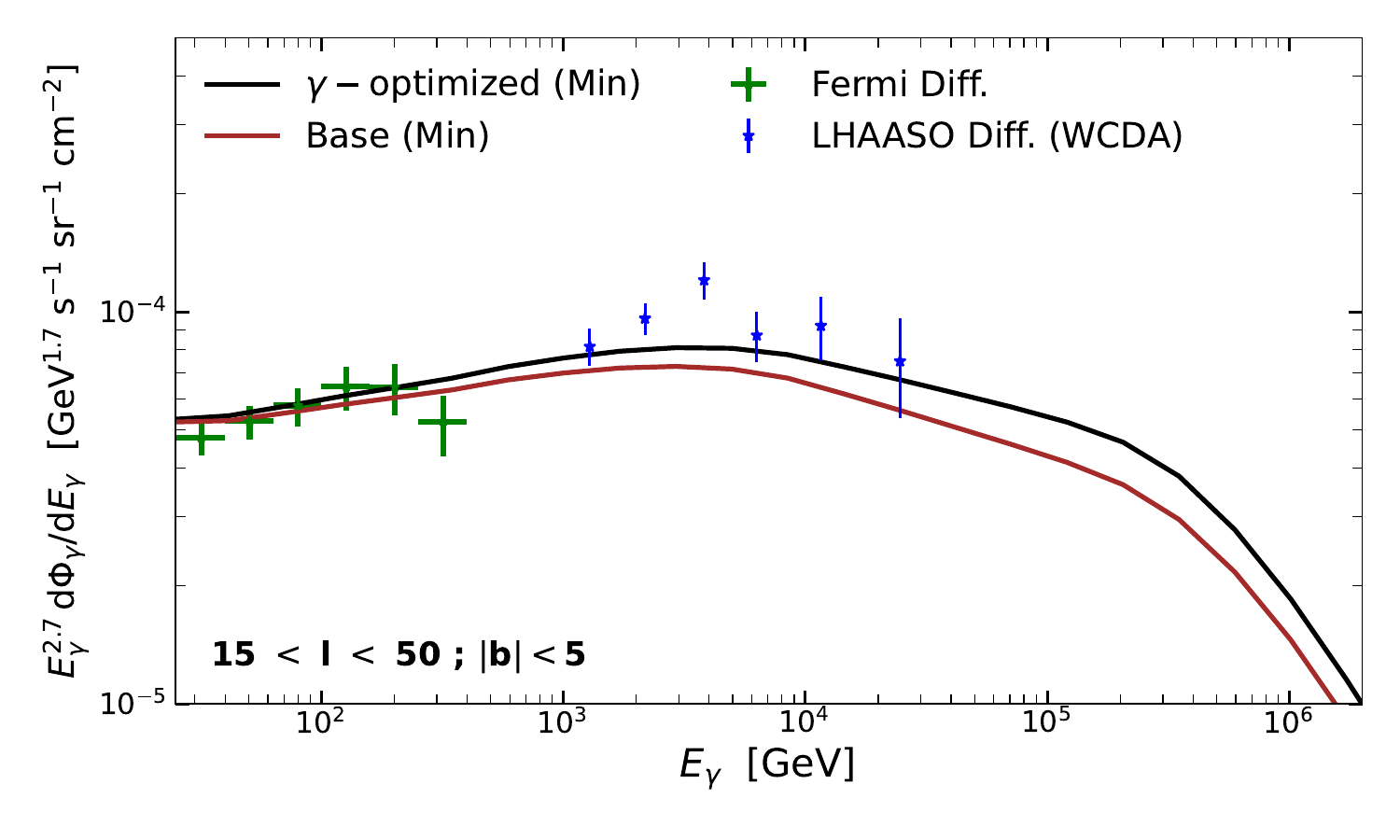}
\caption{The spectra of the Galactic diffuse emission from the $\gamma$-\textit{optimized} and Base models in the Min configuration are compared to the recent diffuse LHAASO-WCDA data (only statistical errors are available) and Fermi-LAT data (with errorbars indicating statistical and systematic uncertainties from the effective area) in the $15^{\circ} < l < 50^{\circ}$, $\vert b \vert < 5^{\circ}$ region. The LHAASO mask is applied to all data sets as well as to the models.}
\label{fig:LHAASO_WCDA_Inner}
\end{figure}

In Fig. \ref{fig:LHAASO_WCDA_Inner} we compare our models with the masked LHAASO-WCDA data.
Again, the LHAASO data looks in better agreement with the spatial dependent models although -- also due to the absence of the, yet not released, systematic errors -- the significance of that preference is still too low to draw robust conclusions.

While one would aspect the offset between conventional and space-dependent models to grow even more with energy, it should taken in mind that LHAASO masks a large fraction the inner GP region which is where the predictions of the two scenarios are most significant.  That effect  is clearly displayed in the left panels of Fig.\ref{fig:Full_LHAASO_Comp}.
A stronger discriminating power is provided by unmasked model and LHAASO data (right panels). That, however, is meaningful only above 100 TeV where the contamination of unresolved sources is expected to be suppressed.
Interestingly, a good agreement -- improving as the energy approaches the PeV -- is found between those data and the $\gamma$-optimized Min model. 


\section{Implications for neutrino astronomy}\label{sec:neutrinos}

In 2023 the IceCube collaboration announced the discovery of the neutrino Galactic diffuse emission ($\nu$-GDE)~\cite{IceCube:2023ame}.
The analysis was based on the statistical comparison of the data obtained from the full-sky cascade event sample with model templates developed to describe Fermi-LAT data up to 10 GeV at least.
The most significant excesses has been obtained following the KRA$_\gamma^5$ model mentioned above and a conventional GALPROP one called $\pi_0$ respectively with $4.37 \sigma$ and $4.71 \sigma$ significance.
Following the spatial and energy distribution of theoretical templates the analysis leaves as a free parameter the normalization of the energy spectrum. 
Clearly, a best-fit normalization too different from that of the models built to reproduce Fermi-LAT data would be problematic looking for a consistent interpretation of $\gamma$-ray data.
Noticeably, the best-fit normalization found by IceCube for the $\pi_0$ set-up is about 4 times larger than that of the original GALPROP model \cite{Fermi-LAT:2016zaq}. 
This issue is likely to be related to the $\pi_0$ model not reproducing the CR hardening found by Fermi-LAT above 10 GeV in the inner GP region \cite{Gaggero:2014xla,Fermi-LAT:2016zaq}.

While the IceCube collaboration found its $\pi_0$ best-fit to be in good agreement with the $\gamma$-GDE measured by Tibet-AS$\gamma$ \cite{TibetASgamma:2021tpz},   
unfortunately, as discussed in the above, the GDE spectrum measured by Tibet is not compatible with the LHAASO results and it is likely to be contaminated by source emissions.
This is especially true taking in mind that conventional models do not predict a large enhancement of the GDE in the GP region masked by LHAASO.
Last but not least the $\pi_0$ model is disfavoured by IceCube results themselves.  In fact, the model independent pre-trial significance map of the all-sky search and the best-fit spectral index angular distribution show the emission to be more concentrated and harder than predicted by that model (see the Supplementary material in~\cite{IceCube:2023ame}).

\begin{figure}[t]
\centering
\includegraphics[width=0.5\textwidth]{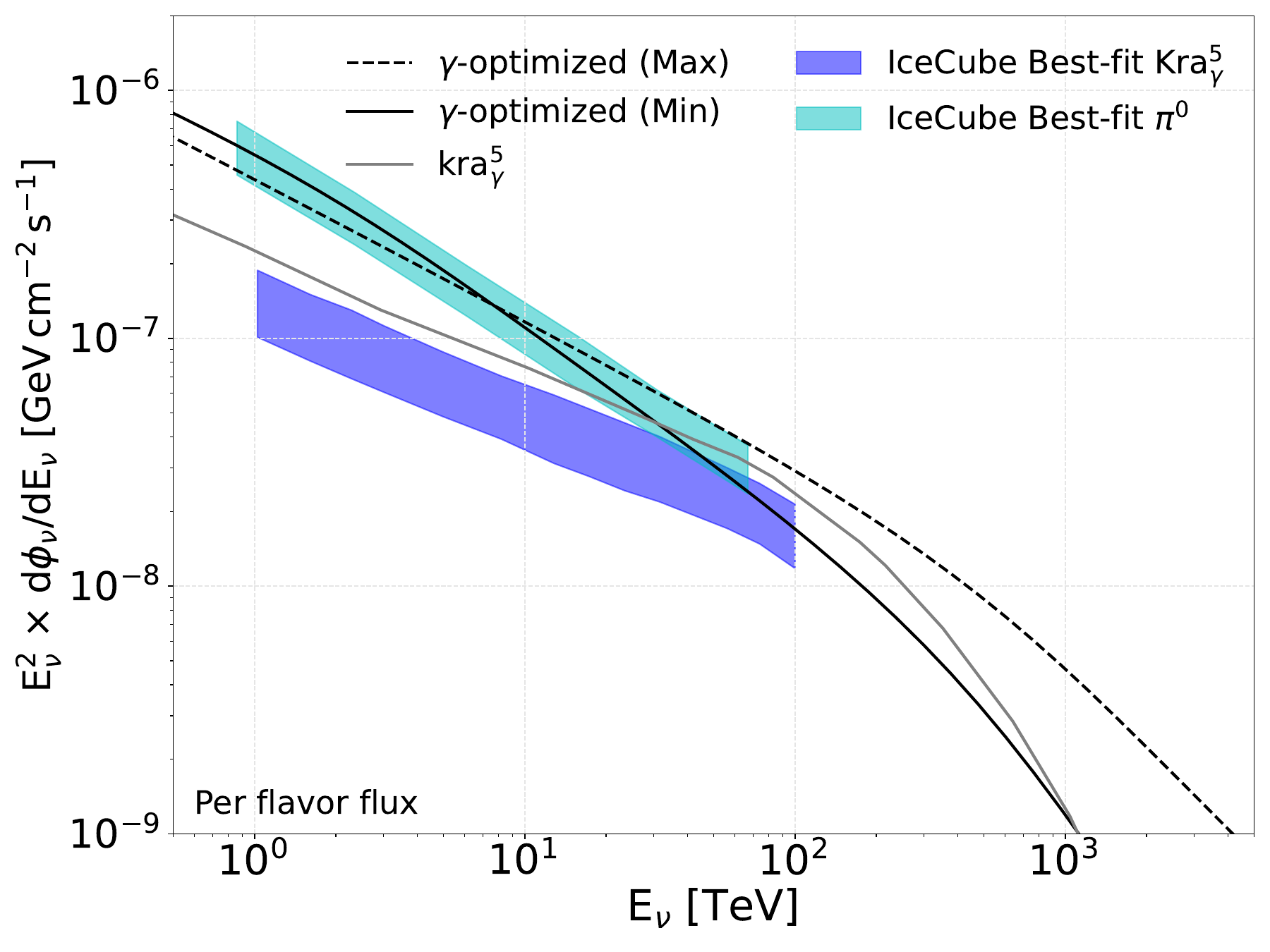}
\caption{The predicted full-sky $\nu$ diffuse emission fluxes (per flavor) from the $\gamma$-\textit{optimized} models are compared to the best-fit IceCube flux ($1 \sigma$ bands) extracted from the KRA-$\gamma$ (cutoff energy of E$_c=5$~PeV) and $\pi^0$ models. The predicted flux from the KRA$_\gamma^5$ model is also reported as a grey line, for comparison. }  
\label{fig:IceCube}
\end{figure}

The KRA$_\gamma^5$ model is in better agreement with those findings.
Also for that model, however, the normalization of the IceCube best-fit is different -- by a factor $0.55$ hence less critically than the $\pi_0$ model -- from that of the original model based on Fermi-LAT data.
We remind to the reader that the KRA$_\gamma^5$ assumes a simplified exponential cutoff of the proton source spectrum at 5 PeV. 
Here we argue that a even better agreement with IceCube results may be obtained using the updated $\gamma$-optimized Min model which, respect to the KRA$_\gamma^5$, adopts a broken power-law source spectrum in order  to match the proton spectrum measured by CALET, DAMPE as well as KASCADE from 10 TeV to 10 PeV.
That choice gives rise to a smaller flux above 10 TeV -- where the signal overcome the atmospheric $\nu$ flux hence giving the largest contribution to the detection significance -- which may take that model closer than the KRA$_\gamma^5$ to a future experimental best-fit.

Waiting for a scrutiny of this expectation -- which would require repeating the IceCube analysis using the $\gamma$-optimized Min model as a template -- may be enlightening to compare our updated models with the best-fit already quoted by IceCube. That is shown in Fig.~\ref{fig:IceCube}.
Interestingly, both $\gamma$-optimized Min and Max setups are consistent with the $\pi_0$ best-fit model, that with the highest significance with respect to IceCube neutrino data. 
The $\gamma$-optimized Min model was also found to be preferred by ANTARES \cite{ANTARES:2022izu,DeLaTorreLuque:2025zsv}, although not with sufficient significance to exclude other scenarios.
A key scrutiny is expected to come from the forthcoming KM3NeT \cite{KM3NeT:2025aps} results.

\section{Conclusions}

We showed that the spectrum and the angular distribution of the $\gamma$-ray diffuse emission of the Galaxy recently measured by the LHAASO collaboration are naturally interpreted in terms of secondary emission by the Galactic CR population if their spectral shape traces that measured by KASCADE (Min setup). 
As long as the mask adopted by LHAASO is used also for Fermi-LAT, both data sets are reproduced by conventional (Base) as well as spatial-dependent ($\gamma$-optimized) CR transport models.
We have shown the degeneracy to be a consequence of that mask cutting the inner GP region where the predictions of those models mostly differ.
Hopefully, in the next future LHAASO may refine its masking procedure so to better cover the inner Galactic plane region removing that degeneracy.    
Interestingly, we found that unmasked LHAASO data already favour the $\gamma$-optimized Min model.
Moreover, it should be stressed that the $\gamma$-optimized models are in better agreement with Fermi-LAT results and should therefore be preferred providing a comprehensive description of the Galactic diffuse emission. 
The SWGO water Cherenkov experiment \cite{SWGO:2025taj}, which will allow a closer view of the GC region, may provide a valuable complementary scrutiny.
 The space-dependent transport scenario seems also to be favoured by IceCube neutrino measurements which found a hint of a spectral hardening in the $\nu$-GDE emission \cite{IceCube:2023ame}.  
While IceCube already found a very good fit ($4.37~\sigma$) using the spatial-dependent KRA$_{\gamma}^5$ model as a template, we argued that agreement may further improve using its $\gamma$-optimized Min update.

\end{document}